\documentclass[aps, 
superscriptaddress,
showpacs,showkeys,pra,amssymb, amsmath]{revtex4-1} 
\usepackage{amsmath,latexsym,amssymb} 
\usepackage[dvips]{graphicx} 
\usepackage{amsmath}
\usepackage{latexsym}
\usepackage{amssymb}  
\usepackage{bm}
 
\begin{document}

\title{Incompleteness in the Bell theorem using 
non-contextual local realistic model}

\author{Koji Nagata} 
\affiliation{Department of Physics, Korea Advanced Institute of Science and Technology, Daejeon 34141, Korea\\
E-mail: $\rm ko_{-}mi_{-}na@yahoo.co.jp$}

\author{Tadao Nakamura}
\affiliation{Department of Information and Computer Science,  
Keio University, 
3-14-1 Hiyoshi, 
Kohoku-ku,
Yokohama 223-8522, 
Japan}

\author{Han Geurdes} 
\affiliation{Geurdes Datascience, KvK 64522202, C vd Lijnstraat 164, 2593 NN, Den Haag\ Netherlands}



\date{\today}

\begin{abstract}
Here, we consider the Bell experiment for a system described by multipartite states in the case where $n$-dichotomic observables are measured per site.
If $n$ is two, we consider a two-setting Bell experiment. 
If $n$ is three, we consider a three-setting Bell experiment.
Two-setting model is an explicit local realistic model for the values of a correlation function, given in a two-setting Bell experiment.
Three-setting model is an explicit local realistic model for the values of a correlation function, given in a three-setting Bell experiment.
In the non-contextual scenario, there is not 
the difference between three-setting model and two-setting model.
And we cannot classify local realistic theories in this case.
This says that we can construct three-setting model from two-setting model.
Surprisingly we can discuss incompleteness in the Bell theorem
using non-contextual model.
On the other hand, 
in the contextual scenario, there is 
the difference between three-setting model and two-setting model.
This says that we must distinguish three-setting model from two-setting model.
And we can classify local realistic theories in this case.

\end{abstract}

\maketitle

\section{Introduction}

Quantum mechanics
(cf. \cite{Neumann,RPF,Redhead,Peres,JJ,NIELSEN_CHUANG}) 
gives accurate and at times remarkably accurate numerical predictions 
for microscopic physical phenomena.

From the incompleteness 
argument of Einstein, Podolsky, and Rosen (EPR) \cite{bib:Einstein}, 
a hidden-variable interpretation
of quantum mechanics
is a topic of 
research \cite{Peres,Redhead}.
One is the Bell-EPR theorem \cite{bib:Bell}. 
This theorem says that the quantum predictions violate the inequality following from the EPR-locality condition.
The locality condition says that a result of measurement pertaining to a system is independent of any measurement performed simultaneously at a distance on another system.
Quantum mechanics does not allow a local realistic interpretation.
Certain quantum predictions violate Bell inequalities \cite{bib:Bell}, which are conditions that a local realistic theory must satisfy.
Experimental efforts (Bell experiment) of a violation of local realism can be seen in \cite{1,2,3}.
Other types of inequalities are given in \cite{other1,other2}.
Bell inequalities with settings other than 
spin polarizations can be seen \cite{otherspin}.

Here, we consider the Bell experiment for a system described by multipartite states in the case where $n$-dichotomic observables are measured per site.
If $n$ is two, we consider a two-setting Bell experiment. 
If $n$ is three, we consider a three-setting Bell experiment.
Two-setting model is an explicit local realistic model for the values of a correlation function, given in a two-setting Bell experiment.
Three-setting model is an explicit local realistic model for the values of a correlation function, given in a three-setting Bell experiment.

Classification of local realistic theories is discussed \cite{NAZU,NAGATAIII}.
We discuss two-setting model cannot construct three-setting model.
The two models are different from each other.
Clearly the argumentations rely on contextual local realistic models.
However, if we take non-contextual local realistic models, 
then the situation changes.
For example,
we may accept Malley's supposition \cite{Malley}, that is,
we suppose all quantum observables in 
a hidden-variable model must commute simultaneously.
This is notable example for non-contextual scenario.

In the non-contextual local realistic model, there is not 
the difference between three-setting model and two-setting model.
This says that we can construct three-setting model from two-setting model.
Surprisingly we can discuss incompleteness (cf.~\cite{GEURDES}) 
in the Bell theorem
by using non-contextual model.

We suppose the Bell theorem as follows \cite{NAGATAIII}:
\begin{eqnarray}
(E_{\rm LR},E)=(E,E).
\end{eqnarray}

In this paper, our aim is of showing
\begin{eqnarray}
(E_{\rm LR},E)<(E,E).
\end{eqnarray}
by using non-contextual local realistic model.  Therefore, 
we discuss incompleteness 
in the Bell theorem by using non-contextual local realistic model. 

On the other hand, 
in the contextual scenario, there must be 
the difference between three-setting model and two-setting model.
This says that we must distinguish three-setting model from two-setting model.
And we can classify local realistic theories in this case.

\section{Incompleteness 
in the Bell theorem using non-contextual local realistic model}

Assume that we have a set of $N$ spins $\frac{1}{2}$. 
Each of them is in a separate laboratory. 
As is well known the measurements (observables) 
for such spins are parameterized by a unit vector $\vec n_j, j=1,2,...,N$ 
(its direction along which the spin component is  measured). 
The results of measurements are $\pm 1$ (in $\hbar/2$ unit). 
We can introduce the ``Bell'' correlation function, 
which is the average of the product of the local results:
\begin{equation}
E(\vec n_1, \vec n_2,\ldots, \vec n_N) 
= 
\langle r_1(\vec n_1) r_2(\vec n_2)\cdots r_N(\vec n_N) \rangle_{\rm avg},
\end{equation}
where $r_j(\vec n_j)$ is the local result, 
$\pm 1$, 
which is obtained if the measurement direction is set at $\vec n_j$.

If an experimental correlation function 
admits rotationally invariant tensor structure
familiar from Newton's theory, 
we can introduce the following form:
\begin{equation}
E(\vec n_1, \vec n_2,\ldots, \vec n_N) 
= 
\hat T \cdot (\vec n_1 \otimes \vec n_2 \otimes \cdots \otimes \vec n_N),
\label{et}
\end{equation}
where $\otimes$ denotes the tensor product, 
$\cdot$ denotes the scalar product in $\mathrm{R}^{\rm 3N}$, 
and $\hat T$ is the correlation tensor given by
\begin{equation}
T_{i_1...i_N}\equiv 
E(\vec x_{1}^{(i_1)},\vec x_{2}^{(i_2)},\ldots, \vec x_{N}^{(i_N)}),
\label{tensor}
\end{equation}
where $\vec x_{j}^{(i_j)}$ is 
a unit directional vector 
of the local coordinate system of the $j$th observer; 
$i_j = 1,2,3$ gives the full set 
of orthogonal vectors 
defining the local Cartesian coordinates. 
Obviously the assumed form of (\ref{et}) 
implies rotational invariance, 
because the correlation function
does not 
depend on 
the coordinate systems used by the observers.
Rotational invariance simply states 
that the value of $E(\vec n_1, \vec n_2,\ldots, \vec n_N)$ cannot depend 
on the local coordinate systems used by the $N$ observers. 

Assume that one knows the values of all $3^N$ components 
of the correlation tensor, $T_{i_1...i_N}$, 
which are obtainable by performing specific $3^N$  measurements 
of the correlation function, (cf. Eq.~(\ref{tensor})). 
Then, with the use of the formula (\ref{et}) 
we can reproduce the values of the correlation functions 
for all other possible sets of local settings.
Using this rotationally invariant structure of the correlation function, 
we shall derive a necessary condition for the existence of 
a local realistic 
theory of the experimental correlation function given in (\ref{et}).
If the correlation 
function is described by the local realistic theory,
then the correlation function must be simulated by the following structure
\begin{eqnarray}
E_{\rm LR}(\vec{n}_1,\vec{n}_2,\ldots,\vec{n}_N)=
\int d\lambda \rho(\lambda)
I^{(1)}(\vec{n}_1,\lambda)I^{(2)}(\vec{n}_2,\lambda)\cdots
I^{(N)}(\vec{n}_N,\lambda),\label{LHVcofun}
\end{eqnarray}
where $\lambda$ is some local hidden variable, 
$\rho(\lambda)$ is a probabilistic distribution, 
and $I^{(j)}(\vec{n}_j,\lambda)$ is 
the predetermined ``hidden'' result of 
the measurement of the dichotomic observable 
$\vec n \cdot \sigma$ with values $\pm 1$.

%%%%%%%%%%%%%%%%%%%%%%%%%%%%%%%%%%%%%%%%%%%%%%%%%%%%%%%%%%%%%%
Let us parametrize the three unit vectors in the plane defined
$\vec{x}^{(1)}_j$ and $\vec{x}^{(2)}_j$ in the following way:
\begin{eqnarray}
\vec{n}_j(\alpha_j^{l_j})
=\cos \alpha_j^{l_j}\vec{x}^{(1)}_j
+
\sin \alpha_j^{l_j}\vec{x}^{(2)}_j,\ \    j=1,2,...,N.
\end{eqnarray}
The phases $\alpha_j^{l_j}$ that experimentalists are allowed to set
are chosen as:
\begin{eqnarray}
\alpha_j^{l_j}=(l_j-1)\pi/3,\ \ l_j=1,2,3.
\end{eqnarray}
We shall show that scalar product of 
the local realistic correlation function
\begin{eqnarray}
E_{\rm LR}(\alpha_1^{l_1},\alpha_2^{l_2},\ldots,\alpha_N^{l_N})=
\int d\lambda \rho(\lambda)
I^{(1)}(\alpha_1^{l_1},\lambda)I^{(2)}(\alpha_2^{l_2},\lambda)\cdots
I^{(N)}(\alpha_N^{l_N},\lambda),\label{LHVANGLE}
\end{eqnarray}
with the rotationally invariant correlation function, that is,
\begin{equation}
E(\alpha_1^{l_1},\alpha_2^{l_2},\ldots,\alpha_N^{l_N})
= 
\hat T \cdot (\vec n_1(\alpha_1^{l_1}) 
\otimes \vec n_2(\alpha_2^{l_2}) \otimes 
\cdots \otimes \vec n_N(\alpha_N^{l_N})),
\end{equation}
is bounded by a specific number dependent on $\hat T$.
We define the scalar product $(E_{\rm LR},E)$ as follows:
We see that the maximal possible value of $(E_{\rm LR},E)$ is bounded as:
\begin{eqnarray}
(E_{\rm LR},E)=\sum_{l_1=1,2,3}\sum_{l_2=1,2,3}\cdots\sum_{l_N=1,2,3}
E_{\rm LR}(\alpha_1^{l_1},\alpha_2^{l_2},\ldots,\alpha_N^{l_N})
E(\alpha_1^{l_1},\alpha_2^{l_2},\ldots,\alpha_N^{l_N})\leq 2^N T_{\rm max},
\label{Bellformula}
\end{eqnarray}
where $T_{\rm max}$ is the maximal possible value 
of the correlation tensor component, i.e.,
\begin{eqnarray}
T_{\rm max}=\max_{\beta_1,\beta_2,...,\beta_N}
E(\beta_1,\beta_2,\ldots,\beta_N),\label{tensormax}
\end{eqnarray}
where $\beta_j$ is some angle.

A necessary condition for 
the existence of the local realistic description $E_{\rm LR}$
of the experimental correlation function
\begin{eqnarray}
E(\alpha_1^{l_1},\alpha_2^{l_2},\ldots,\alpha_N^{l_N})
=E(\vec{n}_1(\alpha_1^{l_1}),
\vec{n}_2(\alpha_2^{l_2}),\ldots,\vec{n}_N(\alpha_N^{l_N}))
\end{eqnarray}
that is for $E_{\rm LR}$ to be equal to $E$ for the three measurement
directions, is that one has $(E_{\rm LR},E)=(E,E)$.
This implies the possibility of modeling $E$ 
by the three-setting local realistic correlation function $E_{\rm LR}$
given in (\ref{LHVANGLE}) with respect to the three measurement directions. 
If we have $(E_{\rm LR},E)<(E,E)$, then the experimental correlation function cannot be explainable by the three-setting local realistic model. (Note that, due to the summation in 
(\ref{Bellformula}), we are looking for the three-setting model.)

In what follows, we derive the upper bound (\ref{Bellformula}).
Since the local realistic model is an average over $\lambda$,
it is enough to find the bound of the following expression:
\begin{eqnarray}
\sum_{l_1=1,2,3}\cdots\sum_{l_N=1,2,3}
I^{(1)}(\alpha_1^{l_1},\lambda)\cdots I^{(N)}(\alpha_N^{l_N},\lambda)
\sum_{i_1,i_2,...,i_N=1,2}
T_{i_1i_2...i_N}c_1^{i_1}c_2^{i_2}\cdots c_N^{i_N},\label{BELLSUM}
\end{eqnarray}
where
\begin{eqnarray}
(c_j^{1},c_j^{2})=(\cos\alpha_j^{l_j},\sin\alpha_j^{l_j}),
\end{eqnarray}
and
\begin{eqnarray}
T_{i_1i_2...i_N}=
\hat T \cdot 
(\vec x_{1}^{(i_1)}\otimes 
\vec x_{2}^{(i_2)}\otimes
\cdots\otimes \vec x_{N}^{(i_N)}),
\end{eqnarray}
compare (\ref{et}) and (\ref{tensor}).

Let us analyze the structure of the sum (\ref{BELLSUM}).
Note that (\ref{BELLSUM}) is a sum, with coefficients given 
by $T_{i_1i_2...i_N}$, which is a product of the following sums:
\begin{eqnarray}
\sum_{l_j=1,2,3}I^{(j)}(\alpha_j^{l_j},\lambda)\cos\alpha_j^{l_j},
\end{eqnarray}
and
\begin{eqnarray}
\sum_{l_j=1,2,3}I^{(j)}(\alpha_j^{l_j},\lambda)\sin\alpha_j^{l_j}.
\end{eqnarray}
We deal here with sums, or rather scalar products 
of $I^{(j)}(\alpha_j^{l_j},\lambda)$ with two-orthogonal vectors.
Ohe has
\begin{eqnarray}
\sum_{l_j=1,2,3}\cos\alpha_j^{l_j}\sin\alpha_j^{l_j}=0,
\end{eqnarray}
because
\begin{eqnarray}
2\times   \sum_{l_j=1,2,3}\cos\alpha_j^{l_j}\sin\alpha_j^{l_j}
=\sum_{l_j=1,2,3}\sin2\alpha_j^{l_j} 
={\rm Im}\left(\sum_{l_j=1,2,3}e^{i2\alpha_j^{l_j}}\right).
\end{eqnarray}
Since $\sum_{l_j=1}^3 e^{i(l_j-1)(2/3)\pi}=0$, the last term vanishes. 

Please note
\begin{eqnarray}
\sum_{l_j=1}^3(\cos\alpha_j^{l_j})^2
=\sum_{l_j=1}^3\frac{1+\cos2\alpha_j^{l_j}}{2}=3/2
\end{eqnarray}
and
\begin{eqnarray}
\sum_{l_j=1}^3(\sin\alpha_j^{l_j})^2
=\sum_{l_j=1}^3\frac{1-\cos2\alpha_j^{l_j}}{2}=3/2,
\end{eqnarray}
because,
\begin{eqnarray}
\sum_{l_j=1,2,3}\cos2\alpha_j^{l_j} 
={\rm Re}\left(\sum_{l_j=1,2,3}e^{i2\alpha_j^{l_j}}\right).
\end{eqnarray}
Since $\sum_{l_j=1}^3 e^{i(l_j-1)(2/3)\pi}=0$, the last term vanishes.

The normalized vectors
$M_1\equiv \sqrt{\frac{2}{3}}
(\cos 0,\cos \pi/3,\cos 2\pi/3)$
and
$M_2\equiv \sqrt{\frac{2}{3}}
(\sin 0,\sin \pi/3,\sin 2\pi/3)$
form a basis of a real two-dimensional plane, which we shall call $S^{(2)}$.
Note further that any vector in $S^{(2)}$ is of the form:
\begin{eqnarray}
A\cdot M_1+B\cdot M_2,
\end{eqnarray}
where $A$ and $B$ are constants, and that any normalized vector in $S^{(2)}$ 
is given by
\begin{eqnarray}
\cos\psi M_1+\sin\psi M_2
=\sqrt{\frac{2}{3}}(\cos (0-\psi),\cos (\pi/3-\psi),\cos (2\pi/3-\psi)).
\end{eqnarray}
The norm $\Vert I^{(j)\Vert} \Vert$ of the projection of $I^{(j)}$ 
into the plane $S^{(2)}$ is given by the maximal possible value of 
the scalar product $I^{(j)}$ with any normalized vector belonging 
to $S^{(2)}$, that is
\begin{eqnarray}
\Vert I^{(j)\Vert} \Vert
=\max_{\psi}\sum_{l_j=1,2,3}
I^{(j)}(\alpha_j^{l_j},\lambda)\sqrt{\frac{2}{3}}
\cos(\alpha_j^{l_j}-\psi)
=\sqrt{\frac{2}{3}}\max_{\psi}{\rm Re}(z\exp(i(-\psi))),
\end{eqnarray}
where $z=\sum_{l_j=1}^3I^{(j)}(\alpha_j^{l_j},\lambda)\exp(i\alpha_j^{l_j})$.
We may assume $|I^{(j)}(\alpha_j^{l_j},\lambda)|=1$.
Then, since $e^{i\alpha_j^{l_j}}=e^{i[(l_j-1)/3]\pi}$,
the possible values for $z$ are 
$0, \pm 2e^{i(\pi/3)}, \pm 2e^{i(2\pi/3)}, \pm 2$.
Note that the minimum possible overall complex phase (modulo $2\pi$)
of $z\exp(i(-\psi))$ is $0$.   Then we obtain
$\Vert I^{(j)\Vert} \Vert\le 
\sqrt{\frac{2}{3}}\times 2\cos 0=2\sqrt{\frac{2}{3}}$.

Since $M_1$ and $M_2$ are two-orthogonal basis vectors in $S^{(2)}$, one has
\begin{eqnarray}
\sum_{l_j=1,2,3}
I^{(j)}(\alpha_j^{l_j},\lambda)\cdot\sqrt{\frac{2}{3}}
\cos \alpha_j^{l_j}=\cos\beta_j\Vert I^{(j)\Vert} \Vert,
\end{eqnarray}
and
\begin{eqnarray}
\sum_{l_j=1,2,3}
I^{(j)}(\alpha_j^{l_j},\lambda)\cdot \sqrt{\frac{2}{3}}
\sin \alpha_j^{l_j}=\sin\beta_j\Vert I^{(j)\Vert} \Vert,
\end{eqnarray}
where $\beta_j$ is some angle.
Using this fact one can put the value 
of (\ref{BELLSUM}) into the following form:
\begin{eqnarray}
\left(\sqrt{\frac{3}{2}}\right)^N \prod_{j=1}^N \Vert I^{(j)\Vert} \Vert
\times \sum_{i_1,i_2,...,i_N=1,2}
T_{i_1i_2...i_N}d_1^{i_1}d_2^{i_2}\cdots d_N^{i_N},\label{Bellsum2}
\end{eqnarray}
where
\begin{eqnarray}
(d_j^1,d_j^2)=(\cos\beta_j,\sin\beta_j).\label{basisvector}
\end{eqnarray}

Let us look at the expression
\begin{eqnarray}
\sum_{i_1,i_2,...,i_N=1,2}
T_{i_1i_2...i_N}d_1^{i_1}d_2^{i_2}\cdots d_N^{i_N}\label{tensorcom}
\end{eqnarray}

Formula (\ref{basisvector}) shows that 
we deal here with two-dimensional unit
vectors $\vec{d}_j=(d_j^1,d_j^2), j=1,2,...,N$, therefore 
(\ref{tensorcom}) is equal to 
$\hat T \cdot (\vec d_1 
\otimes \vec d_2 \otimes 
\cdots \otimes \vec d_N)$, i.e., it is a component of the tensor $\hat T$
in the directions specified by the vectors $\vec{d}_j$.
If one knows all the values of $T_{i_1i_2...i_N}$,
one can always find the maximal possible value of such a component,
and it is equal to $T_{\rm max}$, of equation (\ref{tensormax}).

Therefore since $\Vert I^{(j)\Vert} \Vert\le 2\sqrt{\frac{2}{3}}$
the maximal value of (\ref{Bellsum2}) is $2^N T_{\rm max}$,
and finally one has
\begin{eqnarray}
(E_{\rm LR},E)\leq 2^N T_{\rm max}.\label{FINAL}
\end{eqnarray}

Please note that relation (\ref{FINAL}) is a generalized Bell inequality.
Specific local realistic models, which predict three-setting models,
must satisfy it.
In the following, we shall show that if one replaces
$E_{\rm LR}$ by $E$ one may have a violation of the inequality (\ref{FINAL}).
One has
\begin{eqnarray}
(E,E)=\sum_{l_1=1,2,3}\sum_{l_2=1,2,3}\cdots\sum_{l_N=1,2,3}
\left(\sum_{i_1,i_2,...,i_N=1,2}
T_{i_1i_2...i_N}c_1^{i_1}c_2^{i_2}\cdots c_N^{i_N}\right)^2
=\left(\frac{3}{2}\right)^N\sum_{i_1,i_2,...,i_N=1,2}
T_{i_1i_2...i_N}^2.\label{EE}
\end{eqnarray}
Here, we use the fact that 
$\sum_{l_j=1,2,3}
c_j^{\alpha}c_j^{\beta}
=\frac{3}{2}\delta_{\alpha,\beta}$, because 
$c_j^{1}=\cos\alpha_j^{l_j}$ and
$c_j^{2}=\sin\alpha_j^{l_j}$.

The structure of condition (\ref{FINAL}) and the value (\ref{EE}) suggests that the value of (\ref{EE}) does not have to be smaller than (\ref{FINAL}). That is there may be such correlation functions $E$, which have the property that for any $E_{\rm LR}$ (three-setting model) one has $(E_{\rm LR},E)<(E,E)$, which implies impossibility of modeling $E$ by the three-setting local realistic correlation function $E_{\rm LR}$ with respect to the three measurement directions.

We shall present an important quantum state.
Consider the following
$N$-qubit Greenberger-Horne-Zeilinger (GHZ) state \cite{GHZ}
\begin{equation}
| \psi \rangle = \frac{1}{\sqrt{2}} \Big( 
| z+ \rangle_1 \cdots | z+ \rangle_{N} 
+ | z- \rangle_1 \cdots | z- \rangle_{N} \Big),
\end{equation}
where $| z \pm \rangle_j$ is the eigenstate
of the local $\sigma_z$ operator
of the $j$th observer.
We introduce a mixture
of Greenberger-Horne-Zeilinger correlations
and white noise:
\begin{equation}
\rho = V
|\psi \rangle \langle \psi|
+ (1-V) \rho_{\rm noise},\label{RHO}
\end{equation}
where $|\psi \rangle$ is the 
GHZ state and
$\rho_{\rm noise} = \frac{1}{2^{N}} I$ 
is the random noise admixture. 
The value of $V$ can be 
interpreted as the reduction factor 
of the interferometric contrast 
observed in the $N$-particle correlation experiment.

%%%%%%%%%%%%%%%%%%%%%%%%%%%%%%%%%%%%%%%%%%%%%%%%%%%%%%%%%%%%%%%%%%%%%

Imagine $N$ observers 
who can choose between 
two orthogonal directions 
of spin measurement, 
$\vec x_j^{(1)}$ and $\vec x_j^{(2)}$
for the $j$th one.
Let us assume that the source of $N$ 
entangled spin-carrying particles 
emits them in a state, which can be described as noisy
Greenberger-Horne-Zeilinger correlations, given in (\ref{RHO}).
We can show that 
if the observers limit their settings to 
$\vec x_j^{(1)} = \hat x_j$ and
$\vec x_j^{(2)} = \hat y_j$ 
there are 
\begin{eqnarray}
2^N-1
\end{eqnarray}
components of $\hat T$ of the value $\pm V$. 
These are $T_{11...1}$ and all components 
that except from indices 1 have an even number of indices 2.
Other $x$-$y$ components vanish.

%%%%%%%%%%%%%%%%%%%%%%%%%%%%%%%%%%%%%%%%%%%%%%%%%%%%%%%%%%%%%%%%%%%%%
It is easy to see that $T_{\rm max}=V$ and
$\sum_{i_1,i_2,...,i_N=1,2}
T_{i_1i_2...i_N}^2=V^2 2^{N-1}$.
Then we have $(E_{\rm LR},E)\leq 2^N V$ and
$(E,E)=(\frac{3}{2})^N V^2 2^{N-1}=\frac{3^N}{2}V^2$.
For $N\ge 6$ and $V$ given by
\begin{equation}
 2
\left(\frac{2}{3}\right)^{N}<V \leq \frac{1}{\sqrt{2^{N-1}}}
\end{equation}
we see the fact that 
there exist two-setting local realistic models for 
three measurement directions $(A,B), (B,C), (C,A)$ in consideration 
$\left((0,\frac{\pi}{3},\frac{2\pi}{3})\equiv (A,B,C)  \right)    $
and
these models $(A,B), (B,C), (C,A)$ can construct three-setting 
local realistic models
$(A,B,C)$ because we suppose they are non-contextual local realistic models.
And we have a violation of the Bell theorem:
\begin{eqnarray}
(E_{LR},E)< (E,E).
\end{eqnarray}
That is, non-contextual local realistic models violate
the Bell theorem.

As it is shown in \cite{Zukowski} 
if the correlation tensor satisfies the following conditions
\begin{eqnarray}
&&\sum_{i_1,i_2,\ldots,i_{N}=1,2} T_{i_1i_2...i_{N}}^2 \leq 1,
\label{ZB}
\end{eqnarray}
then there always exists non-contextual local realistic model 
for the set of correlation function values for all directions
lying in a plane.
For our example the condition (\ref{ZB}) 
is met whenever $V \leq \frac{1}{\sqrt{2^{N-1}}}$. 
Nevertheless we have a violation of the Bell theorem
for $V >  2\left(\frac{2}{3}\right)^{N}$.

The situation is such that for $V \leq \frac{1}{\sqrt{2^{N-1}}}$ 
for all two settings per observer
experiments we can construct a local realistic
theory for the values 
of the correlation function 
for the settings 
chosen in the experiment. 
These theories must be consistent with each other
if we want to extend their validity beyond the $2^{N}$  
settings to which each
of them pertains.

Here we suppose they are non-contextual local realistic models.
Then we can extend their validity beyond the $2^{N}$  
settings to which each
of them pertains.
Our calculations clearly indicates that 
this is possible for $V > 2
\left(\frac{2}{3}\right)^{N}$.
And we have the violation of the Bell theorem
for $V >  2\left(\frac{2}{3}\right)^{N}$. 
Therefore, 
we discuss incompleteness 
in the Bell theorem using non-contextual local realistic model.

In the contextual scenario, there is 
the difference between three-setting model and two-setting model.
This says that we must distinguish three-setting model from two-setting model.
And we can classify local realistic theories in this case.

\section{Conclusions}

In conclusions,
in the non-contextual scenario, there has not been
the difference between three-setting model and two-setting model. 
And we cannot have classified local realistic theories in this case.
This has said that we can construct three-setting model from two-setting model.
Surprisingly we can have discussed incompleteness in the Bell theorem
using non-contextual model.
On the other hand, 
in the contextual scenario, there has been 
the difference between three-setting model and two-setting model.
This has said that 
we must distinguish three-setting model from two-setting model.
And we can have classified local realistic theories in this case.

\section*{Note}

On behalf of all authors, the corresponding author 
states that there is no conflict of interest.

\section*{Acknowledgments}

We thank Professor Do Ngoc Diep and
Professor Germano Resconi
for valuable comments.

\end{document}